\documentclass[preprint2]{aastex}

\slugcomment{to appear in ApJ}

\shorttitle{The VSOP 5-GHz AGN Survey}
\shortauthors{Horiuchi et al.}

\begin{document}


\title{The VSOP 5-GHz AGN Survey: \\
IV. The Angular Size/Brightness Temperature Distribution}


\author{S. Horiuchi\footnote{Present address: Swinburne University of
Technology, Hawthorn, Vic. 3122, Australia}} \affil{Jet Propulsion
Laboratory, Pasadena, CA 91109, USA} \email{shoriuchi@swin.edu.au}

\author{E. B. Fomalont}
\affil{National Radio Astronomy Observatory, Charlottesville, VA 22903, USA }
\email{efomalon@nrao.edu}

\author{W. K. Scott \ A.~R.~Taylor}
\affil{University of Calgary, Calgary, Canada}
\email{bill@ras.ucalgary.ca, russ@ras.ucalgary.ca}

\author{J. E. J. Lovell}
\affil{Australia Telescope National Facility, Canberra, Australia}
\email{Jim.Lovell@csiro.au}

\author{G. A. Moellenbrock}
\affil{National Radio Astronomy Observatory, Socorro, NM 87801}
\email{gmoellen@nrao.edu}

\author{R. Dodson, Y. Murata, H. Hirabayashi \& P. G. Edwards}
\affil{Institute of Space and Astronautical Science, Japan Aerospace Exploration Agency Kanagawa 229-8510, Japan}
\email{rdodson@vsop.isas.jaxa.jp, murata@vsop.isas.jaxa.jp,
hirax@vsop.isas.jaxa.jp, pge@vsop.isas.jaxa.jp}

\author{L. I. Gurvits}
\affil{Joint Institute for VLBI in Europe, Dwingeloo, The Netherlands}
\email{lgurvits@jive.nl}

\author{Z-Q. Shen}
\affil{Shanghai Astronomical Observatory, Chinese Academy of Sciences, Shanghai 200030, China}
\email{zshen@shao.ac.cn}





\noindent {\bf ABSTRACT:} The VSOP (VLBI Space Observatory Programme)
mission is a Japanese-led project to study radio sources with
sub-milliarcsec angular resolution, using an orbiting 8-m telescope on
board the satellite HALCA with a global earth-based array of
telescopes.  A major program is the 5 GHz VSOP Survey Program, which
we supplement here with VLBA observations to produce a complete and
flux-density limited sample.  Using statistical methods of analysis of
the observed visibility amplitude versus projected {\em(u,v)} spacing,
we have determined the angular size and brightness temperature
distribution of bright AGN radio cores.  On average, the cores have a
diameter (full-width, half-power) of 0.20 mas which contains about
20\% of the total source emission, and $14\pm 6$\% of the cores are
$<0.04$ mas in size. About $20\pm 5$\% of the radio cores have a
source frame brightness temperature $T_b>1.0\times 10^{13}$K, and
$3\pm 2$\% have $T_b>1.0\times 10^{14}$K.  A model of the high
brightness temperature tail suggests that the radio cores have a
brightness temperatures $\approx 1 \times 10^{12}$ K, and are beamed
toward the observer with an average bulk motion of $\beta=0.993\pm
0.004$.

\section{Introduction}

On 1997 February 12, the Institute of Space and Astronautical Science
of Japan (ISAS) launched a satellite called HALCA, with an 8-m radio
telescope dedicated exclusively to Very Long Baseline Interferometry
(VLBI) \citep{hir98}.  The mission, called VSOP, with a spacecraft
apogee height of 21400 km, gives unparalleled brightness temperature
sensitivity, and allows studies of radio sources with angular
resolution as small as 0.2 mas.  About 75\% of the mission observing
time was devoted to peer-reviewed scientific projects, proposed by the
world-wide astronomical community (called General Observing Time,
GOT).  Many VSOP publications show the complexity and evolution of the
sub-milliarcsec structure of AGNs \citep{pin00,lob01,lis01,tin02,
kam03, mur03,gir04}.

In order to insure that a complete flux-density limited sample of AGNs
were observed during the observing lifetime, the mission-led part, the
VSOP 5-GHz AGN Survey, was given a major portion of the remaining
observing time for sources which were not already included in GOT
proposals.  A general goal of the survey is a compilation of a catalog
of AGN which would be used in part for planning for future space VLBI
missions.  A more immediate goal, reported in this paper, is to
characterize the properties of the sub-milliarcsec structure in AGN,
especially their angular size and brightness temperature
distributions.

   This paper, the fourth in the VSOP AGN Survey series, presents a
non-imaging statistical analysis of the angular size and brightness
temperature distributions of strong AGNs.  This is complementary to
the approach in Paper III \cite{sco04} which shows the images, model
fits, angular sizes and bright temperatures for 102 sources.  A
description of the survey compilation and supporting VLBA observations
was given in Paper I by \cite{hir00}, with additional material in
\cite{fom00a}.  The VSOP observations and data reductions are
described in Paper II by \cite{lov04} (see also \cite{moe00}).  In \S
2, we discuss the source selection and in \S 3 describe how the
observed visibility amplitude versus projected spacing was determined
in order to obtain a statistic which could be used to determine the
angular size properties of AGN.  In \S 4, we derive the angular size
and brightness temperature distributions.  We compare them with other
high resolution surveys and scintillation observations, and fit a
simple model to the distributions.  The major results are summarized
in \S 5.

\section{Compilation of the Data-Base}

\subsection{The VSOP AGN Survey Sample}

The VSOP 5 GHz AGN sample \citep{hir00} was defined to include all
cataloged extragalactic, flat-spectrum radio sources in the sky with
\newline $\bullet$ a total flux density at 5 GHz, $S\geq 0.95$~Jy
\newline $\bullet$ a spectral index $\alpha \geq -0.50$ ($S \propto
\nu^\alpha$) \newline $\bullet$ a galactic latitude $|b|\geq
10^\circ$.
\vskip 0.0cm\noindent This sample contains 344 sources.  This spectral
index criterion eliminated from consideration about 300 extragalactic
sources with flux density $>1$ Jy near the galactic plane, or with
steep radio spectra and thus little milliarcsec structure.  These
sources are arcseconds in size and dominated by double-lobed
structures (eg.~FRI and FRII type radio sources) and are generally
associated with radio galaxies.  At the beginning of 2002 when this
database was assembled, about 50\% of the VSOP observations had been
observed and processed.  Because the selection of observed sources was
randomized by the need to fill in observing holes between the GOT
observations, little bias is introduced by not completing the
observations of the entire list (See Fig.~1 of Paper III) before this
analysis of the data.

    An integral part of the planning for the VSOP Survey was VLBA
observations over a 24-hour period in 1996 June, call the VLBA
pre-launch survey (VLBApls).  All 303 sources of the 344 sample
sources north of declination $-43^\circ$ were observed at 5 GHz, and
the survey results are given by \cite{fom00b}\footnote
{{\it http://www.aoc.nrao.edu/vlba/html/6CM/index.htm}}.  The VLBApls
survey served two important functions for the VSOP Survey.  First, the
VLBA results, with baselines up to 8000 km, indicated which of the AGN
are sufficiently large so that higher resolution VSOP observations on
baselines between 5000 km and 25000 km were not feasible.  Secondly,
when combined with the VSOP data, this enhanced database covers
baselines between 100 km and 25000 km so that a detailed analysis over
a large range of the angular structure of AGNs could be made.
Although the VLBA data were implicitly used in Paper III in order to
obtain VSOP images which were consistent with the lower resolution
data, direct use of the VSOP+VLBA data provides a more accurate
determination of the sub-milliarcsec properties of AGNs, less
susceptible to selection effects, observational biases and the present
incompleteness of the observations.

\subsection {Determining the Visibility Database}

    The observational procedures and data reduction techniques for the
survey experiments are given in Papers II and III.  After calibration
and editing, the basic database for each source is composed of a set
of correlated visibilities (amplitude and phase), measured at many
{\em(u,v)} spacings.  Numerous examples of the character of the data
are given in Fig.~2 of Paper III.  For many of the survey observations
only two or three ground telescopes with HALCA for three or four hours
were used. The large variation of the observation time and ground
telescope participation produced a set of images with a wide range of
quality and resolution, and direct analysis of them can lead to
uncertainties in determining the unbiased properties of AGNs.  Thus,
our statistical analysis uses the visibility data directly in order to
determine the AGN distribution of angular size and brightness
temperature..

     In order to extend the range of resolution of the data, we
concatenated the VSOP data and the VLBApls data, and for this reason
we have restricted the analysis to the 303 survey sources north of
declination $-43^\circ$ which is covered by the VLBA observations.  In
Fig.~1 we give an example of the concatenated data set for J1626$-$2951
and relevant processing.  Fig.~1a shows the visibility amplitude
versus the projected {\em(u,v)} spacing for the VSOP data observed on
22 February 1998, plus the VLBA data observed on 05 June 1996.  The
effect of the source variability is obvious since the VLBA points
which overlap in spacing with the VSOP points are considerably higher.
Using the flux density monitoring of the sources with the Australia
Telescope Compact Array (ATCA) \citep{tin03}, we find that the flux
density of this source in June 1996 was 3.0 Jy, but only 2.1 Jy in
February 1998.  Although the variable component in most AGN are
confined to a small region of the emission extent, the overlap in
spacing and position angle between the VLBA and VSOP surveys are
sufficient to compare their visibility amplitude.  In the example of
Fig.~1a, if the VLBA amplitude scale is multiplied by 2.1/3.0, a
better continuity between the VLBA and VSOP points, as shown in
Fig.~1b, is obtained.  (The large change in flux density for this
source is atypical.)  Source variability corrections to an accuracy of
10\% were obtained from the ATCA and University of Michigan{\footnote{
{\it http.//www.astro.lsa.umich.edu/obs/radiotel/umrao.html}} source
monitoring programs which include most of the sources in the VSOP
sample.  Finally, in order to decrease the size of the database, we
averaged the observed visibility amplitude in bins of width 40
M$\lambda$, and this plot is shown in Fig.~1c.  With the above
processing, we obtained a database containing the visibility
amplitude over a wide range of projected spacing for all sources.

    The VSOP Survey source list of the 303 sources north of
declination $-43^\circ$ is given in Table 1.  It is arranged as
follows:
\begin {description}
\item {} Columns 1 and 2 give the J2000 name, and an alternative name.
\item {} Column 3 lists the total flux density of the source at 5 GHz.
Most of the sources are variable and the total flux density comes from
that of the original finding catalog, as described in \citep{hir00},
or from the VLBA observations.
\item {} Column 4 gives the optical identification: q=quasar, e=elliptical
galaxy, g=faint galaxy ($>24$ mag), b=BLLac object, e=empty field.
\item {} Column 5 gives the redshift, if available.  Many come from
the NASA/IPAC Database (NED, http://nedwww.ipac.caltech.edu)
\item {} Column 6 gives the VSOP observation code.  For additional
information for any source, see the VSOP Survey Program website\footnote{{\it
http://www.jaxa.jp/survey/survey\_db.html}}.
\item {} Column 7 gives the status of the survey observation.  Class~A
(115 sources) means that the VSOP data for the source have been
completely processed and are used in this analysis.  Class~B (124
sources) means that the VSOP data for the source are not yet
available. Class~C (50 sources) are in the VSOP AGN sample, but were
so resolved in the VLBA observations, they were not included as VSOP
targets.  However, they are included as representative of low
brightness objects in the statistical analysis.  Class~D (14 sources)
are so resolved with the VLBApls, that they have been removed from the
VSOP AGN sample as far as the statistical analysis is concerned.
\item {} Column 8 gives the approximate brightness temperature for
sources observed with VSOP.  These are calculated by taking the
correlated flux density at the longest spacing observed for the
source, which is then converted to brightness temperature in units of
$10^{12}$K.  These values are meant solely to be an indication of
which sources have a component of high brightness temperature.  Refer
to Paper III for a more exact determination of the brightness
temperature of source components.  The approximate brightness
temperature here has been used in later discussions.
\item {} Column 9 gives the date of the VSOP observations in day-month-year
format.
\item {} Column 10 gives the data of the VLBA observation in
day-month-year format.  Most were obtained on 05 June 1996 from the
VLBApls survey.  VSOP survey data extracted from the VSOP general
observing time observations which used the VLBA have the same
date as the VSOP observation.
\item {} Column 11 indicates with an 'M' if the source is included in
the 15 GHz Monitoring Of Jets in Active galaxies with VLBA
Experiments (MOJAVE program)\footnote{{\it
http://www.physics.purdue.edu/$\sim$mlister/MOJAVE/}}.
\end{description}

\section {The Relative Visibility versus Projected Spacing}

    The data for all sources were processed to that of the form given
in Fig.~1c.  In order to have a statistic which is independent of the
total flux density of the source, we normalized the measured
correlated flux density of each source to the average visibility
amplitude in the first bin of average spacing 20 M$\lambda$.  We will
denote this normalized flux density as the relative visibility (RV)
and it is a property of the source structure, regardless of the total
intensity of the source.  The properties of the RV with projected
{\em(u,v)} spacing (PS) is the statistic that are used to categorize
the source structure properties of AGN.

    In order to determine an unbiased RV versus PS distribution, we
averaged the contributions of the 303 sources in the following manner.
First, we included the 115 fully-reduced sources, Class=A, 48.1\% of
the sample.  We contend that those sources not yet observed or
reduced, Class=B, have the same average properties of those in
Class=A.  We included a representative portion of the 50 sources
(Class = C) which were too resolved to be observed with HALCA but are
nevertheless part of the VSOP AGN Sample.  In practice we choose all
50, but reduced their weight in the analysis to 0.481 to match the
portion of VSOP observations which have been observed and reduced by
2002 February.

     Fig.~2 shows the dependence of the average RV versus PS for the
sample.  Although data from 165 sources of the 303 in the sample are
included, the distribution should be representative of the entire
sample, as described above.  A preliminary version of this
distribution, with fewer sources, was presented by \cite{lov00}.  The
distribution shows a strong decrease between 20 M$\lambda$ and 140
M$\lambda$, with a less steep decrease at longer spacings.  At 500
M$\lambda$, the RV is 20\% of that at 20 M$\lambda$.  Based on the
measured total flux density of the sources (see Table 1, col 3) the
visibility amplitude at the first point at 20 M$\lambda$ is, on
average, only 52\% of the total source flux density at zero projected
spacing.  Thus, a typical AGN contains about half of its emission in
angular scales $>10$ mas which is invisible in the VLBA and VSOP
observations, but contained in the total source flux density.

     For comparison, the distribution from the VSOP observations of
the Pearson-Readhead (PR) sample at 5 GHz \citep{lis01} is also shown
in Fig.~2.  This sample of 27 sources is defined by $\delta>35^\circ$,
total flux density $>1.3$ Jy, with correlated flux density $>0.4$ Jy
on a 6000 km (100 M$\lambda$) baseline.  Using the same visibility
amplitude normalization as that used with the VSOP Survey sample, the
PR distribution is in reasonable agreement with that found with
the VSOP sample. For the PR sample, there is smaller decrease of the
RV for the longer spacings since the sample definition included a
criterion concerning the compactness of a source.

     The distribution of the RV versus PS can be interpreted as that
produced by a typical AGN structure. We have, thus, fit this
distribution to an average source structure, given by the dashed line
in Fig.~2.  For the space baselines of $PS>180~M\lambda$, the
distribution can be fit with a component of angular size $0.20\pm
0.02$ mas, containing $0.40\pm 0.04$\% of the mas-scale flux density,
or about 20\% of the total flux density.  The fit to shorter spacings
is more ambiguous, and a range of angular scale components are needed.
Most of the emission is contained in an $\approx 1$ mas-sized
component, but some emission is required in a component of $\approx
2.4$ mas to fit the shortest VLBA spacings.  As described above, a
typical AGN contains even more extended emission with a scale size
$>10$ mas which is unobservable with the VLBA or VSOP, but have been
imaged with lower frequency VLBI surveys.

The fit in Fig.~2 of the three components with different angular
scales is consistent with the known structure of many AGNs: The
smallest angular component is the radio core which is generally less
than 1 mas in size.  This component is associated with the inner part
of the radio jet which is often beamed toward the observer. The two larger
components are consistent with the radio jet and internal structures
observed in many sources.  The radio emission, which is completely
resolved out in these observations, is associated with larger-scale
kpc-size emission.  Since the structure of most radio sources is
asymmetric with the radio core at one end of a linear structure, we
originally fit the data in Fig.~2 to an asymmetric spatial
distribution of the three components.  However, the fitting is
insensitive to source asymmetry, but depends strongly on the component
angular scales.

\section {The Statistical Properties of Radio Cores}
\subsection {The Angular Size Distribution}

     Each plotted point in Fig.~2 represents the average RV for the
set of sources.  By fitting this RV distribution to a typical AGN
source model, we obtained an average source structure.  In Fig.~3 we
show the range of values for the RV associated with three spacings; 60
M$\lambda$, 220 M$\lambda$ and 440 M$\lambda$, and the spread of these
values is related to the distribution of the angular size of the AGN
population.  At the shortest spacing of 60 M$\lambda$, about 60\% of
the sources are nearly unresolved (RV$>0.8$), and about 12\% of the
sources have RV$<0.4$.  At the long spacing of 440 M$\lambda$, nearly
40\% of the sources have RV$<0.2$, and only about 30\% have RV$>0.4$.

     In order to determine the range of source sizes which are
consistent with the distributions in Fig.~3, we used the template
structure of the three-component average source model.  In other
words, we assumed that all sources have the same shape RV versus PS,
but are scaled in angular size (or spacing).  We then convolved this
template structure with a two-parameter log-normal distribution
$P(\theta$), where $\theta$ is the angular size of the radio core,
\begin{equation}
P(\theta) d\theta\propto \hbox{exp}
-\Big(\frac{\hbox{log}{(\theta}/\theta_l)}{d}\Bigr)^2 d\theta,
\end{equation}
\noindent with parameters $\theta_l$, the log-mean of the angular size
distribution, and $d$, the dispersion in the log of the angular size.
The best fit for these two parameters was obtained by minimizing the
$\chi^2$ difference between the observed distribution in Fig.~3 with
that expected from the template structure model, convolved with the
distribution in Eq.~(1).  The result of this fit is a core size of
$\theta_l=0.052$ mas and dispersion $d=1.45$, and the fit is shown by
the dark plotted points in Fig.~3.  The error bars represent the
estimated error based on the number of sources used in each spacing
range.

The cross-hatched histogram in Fig.~4 shows the angular size
distribution for this fit.  Approximately 80\% of the radio sources
have a core angular size in the range 0.03 mas to 0.8 mas.  (The two
larger radio components follow the same distribution, but are 5 and 12
times larger than the core.)  About 14\% and 4\% of the sources may
have an angular size less than 0.06 mas and 0.04 mas, respectively.
Although the smaller angular sizes are clearly beyond the resolution
capabilities of these observations of about 0.15 mas for the stronger
cores (Paper III), our assumption of reasonable continuity in the
distribution of angular sizes, implied by the use of Eq. (1), does
infer that these small angular size components are likely to exist.

We also fit the RV vs PS distribution in Fig.~3 with an angular size
model which attempts to minimize the number of small sources.  The
fainter, circle points show a fit to the data with parameters
$\theta_l=0.09$ mas and $d=1.08$.  These values produce an additional
$\chi^2$ deviation from the data which makes this solution (or a more
extreme one) less than 15\% as likely as a solution closer to the best
model.  This angular distribution limit is shown by the open histogram
in Fig.~4, where it is referred to as `largest core'.  For
this distribution, the proportion of sources with cores with an
angular diameter less than 0.06 mas has dropped from 14\% to 6\%, with
only 1\% less than 0.04 mas.  The difference between the two
distributions is an indication of the model error.

\subsection {The Brightness Temperature Distribution}

    The brightness temperature distribution of the core component in
the observer's frame, $T_b\propto S/\theta^2$, can be derived from the
angular size distributions given in Fig.~4, and the average core flux
density, $S$.  To obtain this core flux density, Fig.~5a shows a
similar distribution to that of Fig.~2, but with the y-axis as the
correlated flux density, rather than the RV.  We have divided the
distribution into a low and a high redshift distribution, separated at
$z=0.8$ which is the redshift median value for the sample.  Thus, the
core component, which becomes dominant at spacings greater than 300
M$\lambda$, has an average flux density of 0.5 Jy.  The dependence of
the core flux density with redshift is small, so that the assumption
of converting the angular size distribution to a brightness
temperature distribution using a well-defined average core flux
density of 0.5 Jy is valid.

    The brightness temperature distributions in the observer's
reference frame are shown in Fig.~6, for the best-fit angular size
distribution and the `large-core' angular distribution.  About 14\% of
the sources have $T_b>1.0\times 10^{13}$K for the best fit angular
size distribution; whereas only 6\% are above this temperature for the
large-core fit.  For both distribution, approximately half of all AGN
have a radio core with $T_b>1.0\times 10^{12}$K.  These distributions
agree well with that in Paper III derived from the images. Because our
data modeling does not contain rigid cutoffs at the high resolution
limit of the observations, but uses a reasonable extrapolation below
the formal resolution limit the VSOP observations, the distributions
in Fig.~6 extended to higher brightness temperatures than that from
other VLBI surveys of AGN.  Our addition of AGN which were not
observed with VSOP because of the lack of significant small-scale
structure also extends the brightness temperature distributions to
lower values than other VLBI surveys.

    To correct the brightness temperature $T_b$ from the observer to
the source reference frame, the factor $(1+z)$, where $z$ is the
source redshift, should be applied to the brightness temperature.
Fig.~5b shows the redshift distribution of the 267 sources in the
sample with measured redshifts.  The distribution is relatively flat
out to $z=1.5$, and then drops off with a maximum redshift somewhat
less than 3.0 for this sample.  The average value of $(1+z)$ is 1.81.
The plot in Fig.~5c shows a comparison of the approximate source
brightness temperatures given in Table 1, col (8) versus redshift.
There is clearly little systematic dependence of the brightness
temperature with redshift, hence a simple multiplication of the
brightness temperature scale in Fig.~6 by 1.81 converts from the
observer frame to the source frame.  This assumption should produce an
error no larger than the 15\% difference in the average core flux
density for high and low brightness sources, as shown in Fig.~5a.

    We believe that the statistical analysis of the VSOP+VLBA
observations gives the most realistic and unbiased estimate of the
proportion of high brightness radio cores at 5 GHz yet available.  In
the source reference frame approximately 25\% of the radio cores have
$T_b>1.0\times 10^{13}$K using the best fit angular size distribution,
but this proportion drops to 16\% for the largest core angular size
distribution.  The proportion of the core with $T_b>1.0\times
10^{14}$K is 4\% and 1\% for the two distributions.  This brightness
temperature corresponds to a resolution which is a factor of 10
sharper than the observed VSOP data, but we believe the percentages
are reliable.  Approximately 65\% of the radio cores have
$T_b>1.0\times 10^{12}$K.

    The above results are consistent with other surveys of compact
radio sources.  In the 15-GHz observations of \cite{kov03}, 18 of 160
sources (13\%) have $T_b>1.0\times 10^{13}$K and 46\% have
$T_b>1.0\times 10^{12}$K.  A 22~GHz survey \citep{moe96} found a
similar percentage of high brightness objects.  Our percentage of
cores with $T_b>1.0\times 10^{12}$K is in reasonable agreement (as it
should be since the data overlap is large) with the 53\% found by
\cite{sco04} from the images and model-fitting of 102 sources from the
VSOP Survey\footnote{Twenty sources in the Scott sample are not
included in this statistical analysis because they were south of
$\delta<-43$, or not within the strict definition of the catalog
completeness.  About 30 additional sources, with good visibility data,
but not yet imaged, are included this statistical analysis.}.  The
highest brightness temperatures that has been measured for an
individual source is $5.8\times 10^{13}$K for AO0235+164 at 5 GHz
\citep{fre00}.

    The scintillation of many radio sources also implies that they
contain radio components with $T_b>10^{14}$ K \citep{ked01}.  Of the
710 sources surveyed with the VLA at 4.9 GHz, the Micro-Arcsecond
Scintillation-Induced Variability Survey (MASIV) \citep{lov03}, about
12\% were classified as variable.  The variability in the total flux
density averaged about 6\%, and that the variable time scales varied
from a few hours to a few days.  An approximate interpretation of
these results is: about 12\% of the AGNs contain a radio component
with 6\% of the total source flux density and $T_b\ge10^{14}$ K.  The
results from our statistical analysis suggests that about 20\% of the
AGNs contain a radio component with 20\% of the total source flux
density and $T_b\ge 10^{13}$K.  These two descriptions of the
properties of the high brightness radio cores are compatible.
Finally, a correlation between sources which scintillate and those
which have relatively large correlated flux density observed by VSOP
was reported by \cite{lis01b} using the PR sample of sources. The
detailed study of the scintillation properties of the VSOP sample is
in progress.

\subsection {Brightness Temperature Modeling}

It is generally assumed that the maximum brightness temperature from a
synchrotron emitting radio source is $T_{max}\approx 10^{12}$K,
because the strong inverse-Compton emission will quickly quench
the radio emission above this brightness
\citep{kel69,lis01,tin01,kel02}.  Calculations based on equipartion of
energy arguments suggest that this limit may be as low as $10^{10.5}$ K
\citep{rea96}. Other emission mechanisms have also been proposed,
including relativistic induced Compton scattering \citep{sin94} and
coherent synchrotron emission processes \citep{mel02}, which have a
higher brightness temperature limit.  The observed brightness
temperatures above $T_{max}$, however, can be produced by the Doppler
boosting {\citep{shk63} of the emission from the radio core material
which moves with a relativistic bulk velocity $v_b$ nearly in the
direction to the observer, given by the angle, $\psi$.

In order to obtain estimates of the three parameters, $T_{max},
\beta=v_b/c$ and $\psi$ needed to reproduce the high brightness tail
of the distribution in Fig.~6, we used the following simple model: The
number density of the source brightness temperature, $T<T_{max}$, is
proportional to $\log(T_{max}/T)$; the motion of this material is
$\beta$, and the maximum orientation to the line of sight is $\psi$.
For a more detailed modeling of superluminal sources, see
\cite{ver94}.  The plotted points in Fig.~6 are those for the values
$T_{max} = 1.0\times 10^{12}$K, $\beta=0.993$, $\psi = 35^\circ$.
Reasonable fits are obtained for the range $0.5 < T_{max} < 2.0$ and
$0.990 < \beta < 0.997$.  Similar model parameters are obtained from
multi-epoch observations of the \lq superluminal\rq~motions of radio
components, which suggest $\beta\approx 0.99$, corresponding to
$\gamma=(1-\beta^2)^{-0.5}\approx 10$ \citep{ver94,kel00,kov03}.

\section {Conclusions}

     Before the launch of HALCA, there was little {\it direct}
evidence concerning the brightness temperature distribution of radio
components associated with AGN.  Figs.~2 and 3 show that at 5 GHz with
baselines up to 25,000 km there is significant emission for many AGNs,
and future use of a space-VLBI mission with substantially longer
baselines are required to probe the evolution and structure of these
high brightness radio cores.  Future space VLBI missions with longer
baselines and substantially improved sensitivity are, thus, required
to probe the evolution and structure of these high brightness cores in
AGNs.

    The VSOP AGN Survey was compiled in order to determine the
properties of the sub-mas radio structure of strong AGN.  The source
sample covered the entire sky ($|b|>10^\circ$) and included sources at
5 GHz above 1 Jy, with a relatively flat spectral index.  No other
criteria were used.  The analysis in this paper provides an attempt at
an unbiased determination of the radio source structure parameters by
using the measured RV versus PS properties in a simple and
straight-forward manner.  An analysis from the derived images and
models of the individual sources \citep{sco04} obtains similar results
on the angular size and brightness temperature of the AGNs.

    The major properties of AGN derived from the statistical analysis
described in this paper are:
\begin{itemize}
\item About 50\% of the total emission from an average AGN is
completely resolved using at the shortest VLBA spacings, and are
therefore contained in a component $>10$ mas in size.
\item About 40\% of the milliarcsec emission (20\% of the total
emission) comes from a radio core of average size $0.20\pm 0.02$ mas.
\item About 80\% of the radio cores have an angular size in the range
of 0.03 to 0.8 mas.  We estimate that $10\pm 4$\% of the cores are
$<0.06$ mas at 5 GHz.
\item A majority of the AGN radio cores have a brightness temperature in
excess of $1.0\times 10^{12}$ K, and we estimate that $20\pm 5$\% of
the cores have $T_b>1.0\times 10^{13}$K and $3\pm 2$\% have
$T_b>1.0\times 10^{14}$K in the source reference frame.
\item The derived brightness temperature distribution is in good
agreement with the results from other high-resolution radio source
surveys and with radio scintillation observations.
\end{itemize}

\acknowledgments

We gratefully acknowledge the VSOP project, which is led by the
Institute of Space and Astronautical Science (ISAS, now called the
Japan Aerospace Exploration Agency, JAXA) in cooperation with many
organizations and radio telescopes around the world.  The National
Radio Astronomy Observatory is a facility of the National Science
Foundation, operated under cooperative agreement by Associated
Universities, Inc. SH acknowledges support through an NRC/NASA-JPL
Research Associateship: WKS thanks the support from the Canadian Space
Agency: RD is supported by the Japanese Society for the Promotion of
Science: JEJL thanks the support from the Australian Commonwealth
Scientific \& Industrial Research Organisation.  This research has
made use of the NASA/IPAC Extragalactic Database (NED) which is
operated by the Jet Propulsion Laboratory, California Institute of
technology, under contract with the National Aeronautics and Space
Administration.  Finally, we thank an industrious referee for major
improvements to the organization and content of the paper.

\clearpage


\begin{figure}
\includegraphics{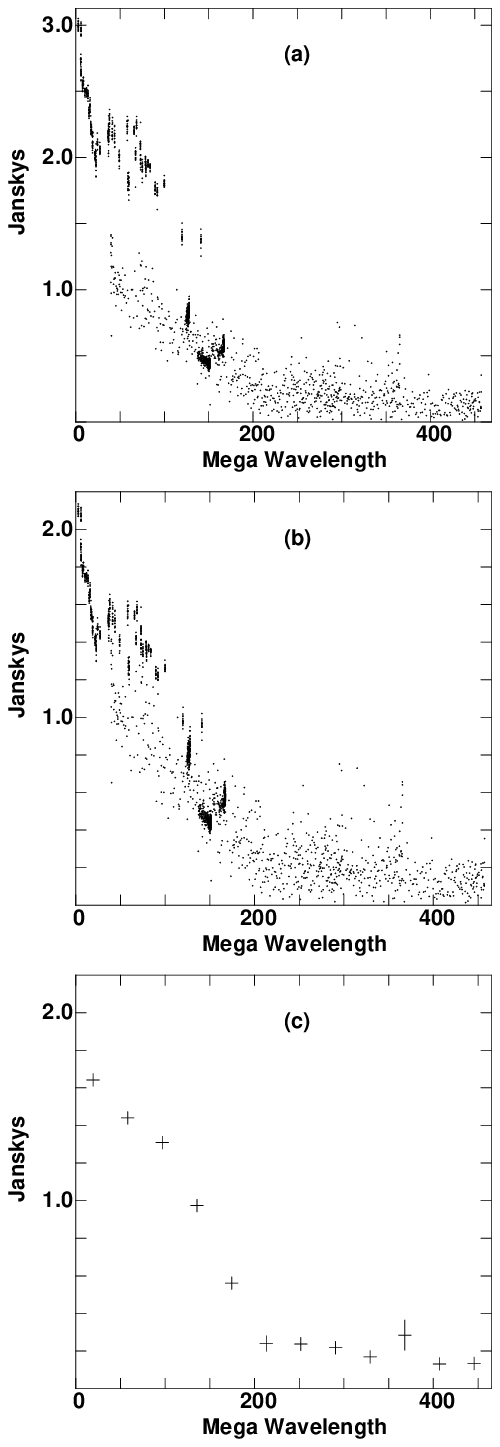}
\vspace{10cm}
\caption{{\bf Combining and Averaging the VSOP and VLBA Data:} (a):
The visibility amplitude versus the projected spacing {\em(u,v)} is
shown for the source J1626$-$2951.  The VSOP data at the longer spacings
are shown by the fainter dots; the VLBA data at the shorter spacings
are shown by the darker dots.  These observations were taken nearly
two years apart.  (b): The same plot as in (a) after correcting the
VLBA amplitude scale, based on the measured variability of the
source. (c): Each point shows the average visibility amplitude and
error estimate for all data within each interval of 40 M$\lambda$
projected spacing.}
\end{figure}
\newpage

\begin{figure}
\includegraphics{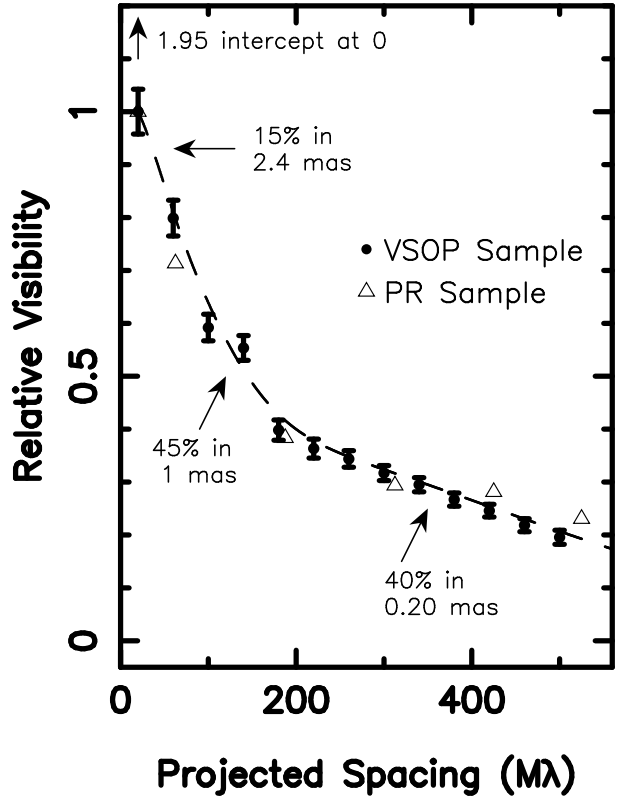}
\vspace{8cm}
\caption{{\bf The Relative Visibility versus Projected Spacing for the
VSOP AGN Sample:} The relative visibility distribution, normalized to
1.0 at 20 M$\lambda$, is shown by the plotted solid points.  The error
estimate is based on the number of sources compiled for each point.
The distribution for the Pearson-Readhead sample of 24 sources is
given by the open triangles.  The dashed line shows the dependence of
the relative visibility for a three component model with component
sizes and proportions as indicated.  Each spacing bin is 40 M$\lambda$
wide and not all bins were sampled for all sources.  The maximum
resolution for a source observed with VSOP varied between 300 to 540
M$\lambda$, the maximum resolution for VLBA observations was 150
M$\lambda$.  At zero spacing, the average visibility is 1.95 times
that at 20 M$\lambda$.  A spacing of 500 M$\lambda$ corresponds to
25,000 km.}
\end{figure}
\newpage

\begin{figure} \includegraphics{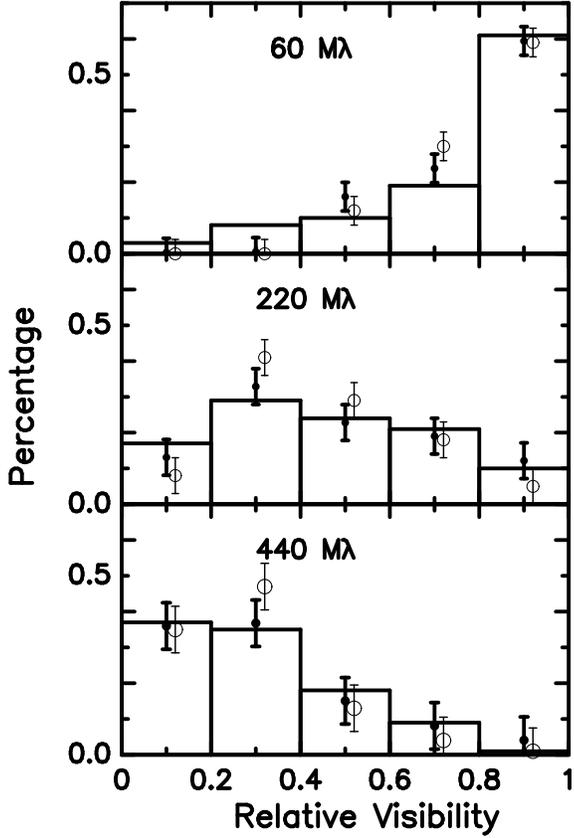} 

\vspace{10cm} \caption{{\bf The Fit to the Relative Visibility
Distributions Versus Spacing:} The histogram of the relative
visibility for the sources in the sample are given for the three
indicated spacings.  The plotted points show the distribution from the
template source model convolved with a core angular-size distribution
shown in Fig.~3, using the parameters of Eq.(1).  The error bars are
estimated from the number of observations at each resolution.  The
lighter plotted points show a relatively poor fit associated with a
model with a larger core diameter (see text).}
\end{figure}

\begin{figure}
\includegraphics{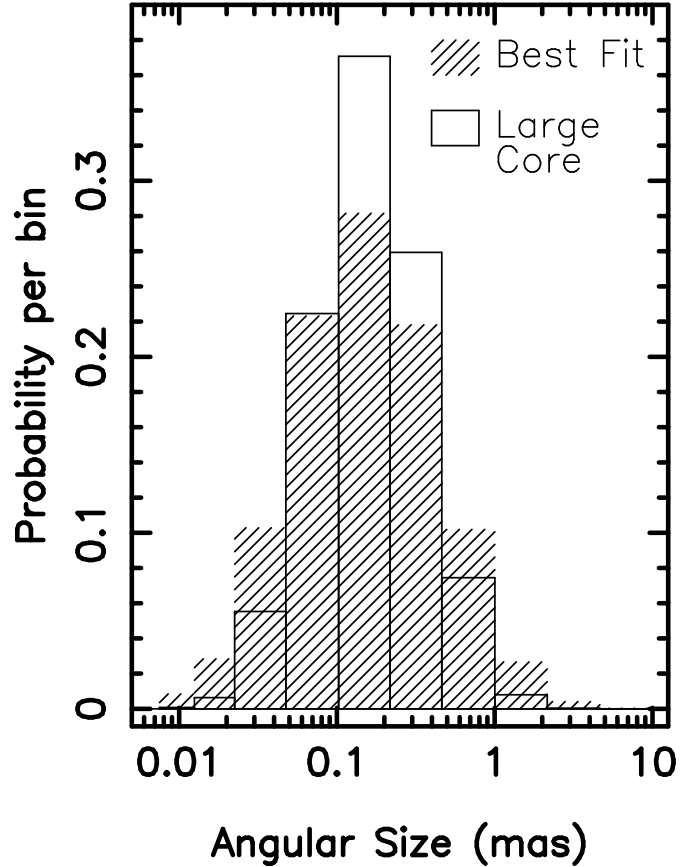}
\vspace{10cm}
\caption{{\bf The Derived Core Angular-Size Distributions:} The
probability distribution of the core angular size distribution with
$\theta_l=0.052$ mas and $d=1.45$ (see Eq.~(1)) is shown by the
cross-hatched histogram, correspond to the darker plotted points in
Fig.~3.  The open histogram shows the distribution associated with the
model for a larger core, $\theta_l=0.09$ mas and $d=1.08$, shown by
the lighter plotted points in Fig.~3.}
\end{figure}

\begin{figure}
\includegraphics{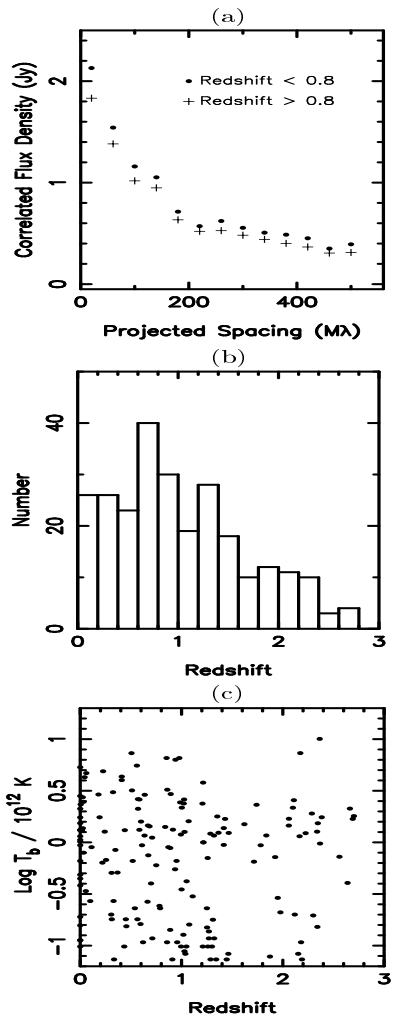}
\vspace{12cm}
\caption{{\bf (a) The Correlated Flux Density versus Projected Spacing
for Sources in the Sample:} This plot is similar to Fig.~2, but with
the ordinate values in correlated flux density, rather than normalized
to 1.0 at $20~M\lambda$.  The sample was split into two nearly equal
parts, defined by the source redshift.  The higher redshift sample has
about 15\% less flux density at all spacings than the lower redshift
sample. {\bf (b) The Distribution of redshift:} The number
distribution for the 277 of the 303 sources in the sample with a
measured redshift. {\bf (c) The Dependence of the Brightness
Temperature with Redshift:} The approximate brightness temperatures,
given in Table 1, Col (8) is plotted versus redshift if available.
Little core flux density versus redshift is observe; hence, the
correction from the observer's frame to the source frame by simply
increasing the brightness temperature scale by 1.81 is valid.}
\end{figure}

\begin{figure}
\includegraphics{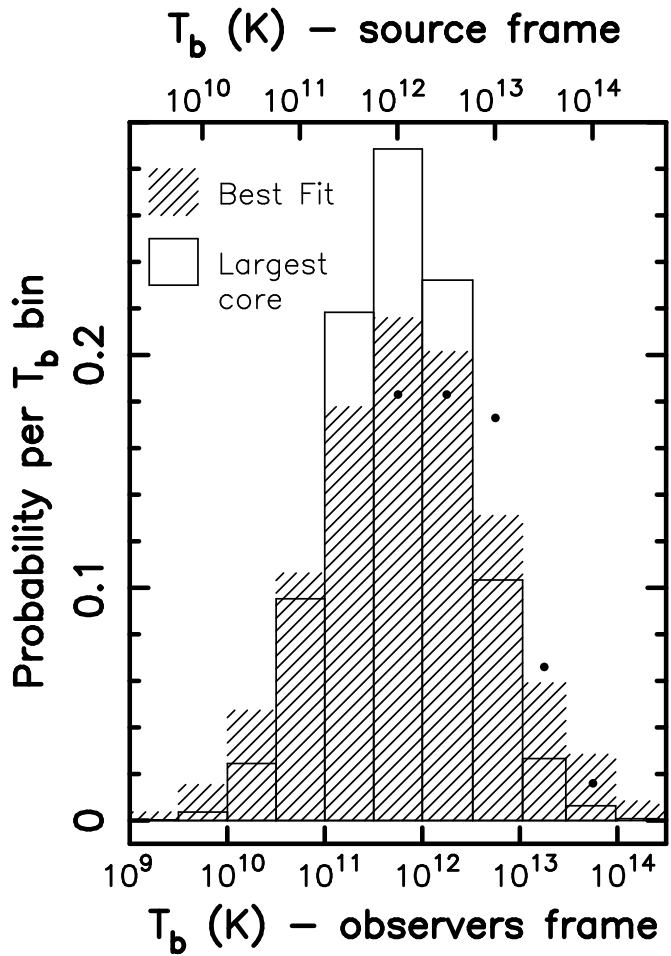}
\vspace{12cm}
\caption{{\bf The Brightness Temperature Distributions of the VSOP AGN
Sample:} The brightness temperature distribution, implied by the
core-angular size distributions, shown in Fig.~3.  The cross-hatched
distribution is the best fit to the VSOP AGN sample; the open
distribution is for the model with the larger core size.  The plotted
points at the high brightness temperature tails are shown for the
model discussed in the text, with $v_b/c=0.993$, $T_{max}=1\times
10^{12}$K, and $\psi=35^\circ$.}
\end{figure}

\clearpage

\begin{deluxetable}{llrclccrccc}
\tablecolumns{11}
\tablecaption{VSOP AGN Survey Source List}
\tabletypesize{\footnotesize}
\tablehead {

\multicolumn{1}{c} {(1)} &
\multicolumn{1}{c} {(2)} &
\multicolumn{1}{c} {(3)} &
\multicolumn{1}{c} {(4)} &
\multicolumn{1}{c} {(5)} &
\multicolumn{1}{c} {(6)} &
\multicolumn{1}{c} {(7)} &
\multicolumn{1}{c} {(8)} &
\multicolumn{1}{c} {(9)} &
\multicolumn{1}{c} {(10)} &
\multicolumn{1}{c} {(11)} \\
\multicolumn{1}{c} {IAU Name} &
\multicolumn{1}{c} {Alt Name} &
\multicolumn{1}{c} {S$_5$} &
\multicolumn{1}{c} {ID} &
\multicolumn{1}{c} {z} &
\multicolumn{1}{c} {Exp} &
\multicolumn{1}{c} {Class} &
\multicolumn{1}{c} {T$_b$} &
\multicolumn{1}{c} {OBS$_{vsop}$} &
\multicolumn{1}{c} {OBS$_{vlba}$} &
\multicolumn{1}{c} {MOJAVE} \\

}
\startdata
J0006$-$0623 & 0003$-$066 &   2.5 & g & 0.347  &  vs03t & A &  6.5 & 29-08-98 & 05-06-96 & M\\
J0013$+$4051 & 0010$+$405 &   1.0 & g & 0.255  &  vs13a & B &      &          & 05-06-96 &  \\
J0019$+$7327 & 0016$+$731 &   1.7 & q & 1.781  &  vs07a & A &  1.9 & 02-03-98 & 02-03-98 & M\\
J0029$+$3456 & 0025$+$345 &   1.3 & g & 0.600  &        & C &      &          & 05-06-96 & M\\
J0038$+$4137 & 0035$+$413 &   1.1 & q & 1.353  &        & C &      &          & 05-06-96 & M\\
J0042$+$2320 & 0039$+$230 &   1.6 & e &        &  vs07u & A &  1.3 & 12-08-99 & 05-06-96 &  \\
J0105$+$4819 & 0102$+$480 &   1.1 & e &        &  vs08q & B &      &          & 05-06-96 &  \\
J0106$-$4034 & 0104$-$408 &   2.6 & q & 0.584  &  vs03s & A &  6.3 & 08-06-98 & 05-06-96 &  \\
J0108$+$0135 & 0106$+$013 &   3.4 & q & 2.099  &  vs02t & B &      &          & 05-06-96 & M\\
J0111$+$3906 & 0108$+$388 &   1.3 & g & 0.669  &        & C &      &          & 05-06-96 & M\\
J0115$-$0127 & 0112$-$017 &   1.6 & q & 1.365  &  vs07t & A &  1.4 & 10-08-99 & 05-06-96 & M\\
J0116$-$1136 & 0113$-$118 &   1.9 & q & 0.672  &  vs06h & A &  1.7 & 15-07-01 & 05-06-96 &  \\
J0119$+$0829 & 0116$+$082 &   1.2 & g & 0.584  &        & C &      &          & 05-06-96 &  \\
J0119$+$3210 & 0116$+$319 &   1.6 & g & 0.060  &        & D &      &          & 05-06-96 &  \\
J0120$-$2701 & 0118$-$272 &   1.2 & b & 0.559  &        & C &      &          & 05-06-96 &  \\
J0121$+$1149 & 0119$+$115 &   1.1 & q & 0.570  &  vs08p & A &  0.1 & 22-07-01 & 05-06-96 & M\\
J0121$+$0422 & 0119$+$041 &   2.0 & q & 0.637  &  vs05y & A &  0.2 & 01-08-98 & 05-06-96 & M\\
J0125$-$0005 & 0122$-$003 &   1.6 & q & 1.070  &  vs05x & A &  1.3 & 02-08-98 & 05-06-96 &  \\
J0126$-$0123 & 0123$-$106 &   1.3 & e &        &        & D &      &          & 05-06-96 &  \\
J0126$+$2559 & 0123$+$257 &   1.4 & q & 2.364  &  vs08o & A &  0.8 & 20-07-01 & 05-06-96 &  \\
J0132$-$1654 & 0130$-$171 &   1.2 & q & 1.022  &  vs08c & B &      &          & 05-06-96 &  \\
J0136$+$4751 & 0133$+$476 &   2.4 & q & 0.859  &  vs03r & A &  3.7 & 16-08-99 & 16-08-99 & M\\
J0141$-$0928 & 0138$-$097 &   1.2 & b & 0.440  &  vs10i & B &      &          & 05-06-96 &  \\
J0149$+$0555 & 0146$+$056 &   1.5 & q & 2.345  &  vs06z & A &  1.7 & 27-07-01 & 05-06-96 &  \\
J0152$+$2207 & 0149$+$218 &   1.1 & q & 1.320  &  vs08n & B &      &          & 05-06-96 & M\\
J0153$-$3310 & 0150$-$334 &   1.4 & q & 0.610  &        & C &      &          & 05-06-96 &  \\
J0155$-$4048 & 0153$-$410 &   1.2 & g & 0.226  &        & C &      &          & 05-06-96 &  \\
J0157$+$7442 & 0153$+$744 &   1.5 & q & 2.338  &        & C &      &          & 05-06-96 & M\\
J0204$+$1514 & 0202$+$149 &   2.7 & q &        &  vs03i & B &      &          & 05-06-96 & M\\
J0204$-$1701 & 0202$-$172 &   1.4 & q & 1.740  &  vs08b & B &      &          & 05-06-96 &  \\
J0205$+$3212 & 0202$+$319 &   1.0 & q & 1.466  &  vs12c & B &      &          & 05-06-96 & M\\
J0217$+$7349 & 0212$+$735 &   3.2 & b & 2.367  &  vs02o & A &  1.5 & 05-09-97 & 05-09-97 & M\\
J0217$+$0144 & 0215$+$015 &   1.6 & b & 1.715  &  vs07r & A &  0.1 & 03-08-98 & 05-06-96 &  \\
J0221$+$3556 & 0218$+$357 &   1.5 & q & 0.936  &        & C &      &          & 05-06-96 & M\\
J0224$+$0659 & 0221$+$067 &   1.0 & q & 0.511  &  vs12b & B &      &          & 05-06-96 & M\\
J0226$+$3421 & 0223$+$341 &   1.7 & q &        &  vs05o & B &      &          & 05-06-96 &  \\
J0231$+$1322 & 0229$+$131 &   2.4 & q & 2.059  &  vs04l & A &  2.1 & 13-08-99 & 05-06-96 &  \\
J0237$+$2848 & 0234$+$285 &   3.4 & q & 1.213  &  vs02s & B &      &          & 05-06-96 & M\\
J0238$+$1636 & 0235$+$164 &   1.9 & b & 0.940  &        & C &      &          & 05-06-96 & M\\
J0239$+$0416 & 0237$+$040 &   1.1 & q & 0.978  &  vs11i & B &      &          & 05-06-96 &  \\
J0240$-$2309 & 0237$-$233 &   3.6 & q & 2.225  &        & C &      &          & 05-06-96 & M\\
J0241$-$0815 & NGC1052    &   3.2 & g & 0.0049 &        & C &      &          & 05-06-96 &  \\
J0242$+$1101 & 0239$+$108 &   1.6 & q &        &  vs05w & B &      &          & 05-06-96 &  \\
J0251$+$4315 & 0248$+$430 &   1.4 & q & 1.310  &  vs07q & A &  1.1 & 15-02-99 & 15-02-99 &  \\
J0259$+$0747 & 0256$+$075 &   1.0 & b & 0.893  &  vs12a & B &      &          & 05-06-96 &  \\
J0312$+$0133 & 0310$+$013 &   1.0 & q & 0.664  &  vs11z & B &      &          & 05-06-96 & M\\
J0318$+$4151 & NGC1265    &   1.5 & g & 0.025  &        & D &      &          & 05-06-96 &  \\
J0319$+$4130 & 3C84       &  42.4 & g & 0.0175 &  vs01c & A &  1.2 & 25-08-98 & 25-08-98 &  \\
J0321$+$1221 & 0319$+$121 &   1.6 & q & 2.662  &  vs07p & A &  2.1 & 14-02-01 & 05-06-96 &  \\
J0336$+$3218 & NRAO140    &   2.0 & q & 1.258  &  vs05f & B &      &          & 05-06-96 & M\\
J0339$-$0146 & CTA26      &   3.0 & q & 0.852  &  vs03q & A &  1.4 & 17-02-02 & 05-06-96 & M\\
J0348$-$2749 & 0346$-$279 &   1.4 & q & 0.987  &  vs08m & A &  1.3 & 09-08-01 & 05-06-96 &  \\
J0402$-$3147 & 0400$-$319 &   1.0 & q &        &  vs11y & B &      &          & 05-06-96 &  \\
J0403$+$2600 & 0400$+$258 &   1.0 & q & 2.109  &  vs11x & B &      &          & 05-06-96 &  \\
J0403$-$3605 & 0402$-$362 &   2.2 & q & 1.417  &  vs03z & A &  0.1 & 30-07-98 & 05-06-96 &  \\
J0405$-$1308 & 0403$-$132 &   3.2 & q & 0.571  &  vs03e & A &  2.3 & 19-08-98 & 05-06-96 &  \\
J0406$-$3826 & 0405$-$385 &   1.3 & q & 1.285  &  vs07o & B &      &          & 15-09-99 & M\\
J0412$+$2305 & 0409$+$229 &   1.0 & q & 1.215  &        & C &      &          & 05-06-96 &  \\
J0414$+$3418 & 0411$+$341 &   1.6 & e &        &        & C &      &          & 05-06-96 &  \\
J0414$+$0534 & 0411$+$054 &   1.1 & q & 2.639  &        & C &      &          & 05-06-96 &  \\
J0423$-$0120 & 0420$-$014 &   4.4 & q & 0.915  &  vs02g & A &  2.3 & 04-02-99 & 04-02-99 & M\\
J0424$-$3756 & 0422$-$380 &   1.7 & q & 0.782  &  vs06y & B &      &          & 05-06-96 &  \\
J0424$+$0036 & 0422$+$004 &   1.5 & b & 0.310  &  vs07z & A &  2.6 & 24-08-99 & 05-06-96 &  \\
J0428$-$3756 & 0426$-$380 &   1.8 & b & 1.030  &  vs05e & B &      &          & 05-06-96 &  \\
J0431$+$2037 & 0428$+$205 &   2.8 & g & 0.219  &        & C &      &          & 05-06-96 &  \\
J0433$+$0521 & 3C120      &   3.8 & g & 0.0329 &  vs02n & A &  1.3 & 11-02-99 & 11/02/99 & M\\
J0437$-$1844 & 0434$-$188 &   1.2 & q & 2.702  &  vs09m & B &      &          & 05-06-96 &  \\
J0449$+$1121 & 0446$+$112 &   1.0 & g & 1.207  &  vs10h & B &      &          & 05-06-96 &  \\
J0453$-$2807 & 0451$-$282 &   2.5 & q & 2.559  &  vs04g & A &  4.0 & 12-09-99 & 05-06-96 &  \\
J0455$-$3006 & 0453$-$301 &   1.5 & g &        &        & D &      &          & 05-06-96 &  \\
J0457$-$2324 & 0454$-$234 &   2.0 & b & 1.003  &  vs05n & A &  1.5 & 27-02-02 & 05-06-96 &  \\
J0459$+$0229 & 0457$+$024 &   1.7 & q & 2.384  &  vs05m & B &      &          & 05-06-96 &  \\
J0501$-$0159 & 0458$-$020 &   3.3 & q & 2.286  &  vs02z & A &  1.2 & 22-09-99 & 05-06-96 & M\\
J0503$+$0203 & 0500$+$019 &   2.1 & q & 1.000  &  vs04a & B &      &          & 05-06-96 &  \\
J0508$+$8432 & 0454$+$844 &   1.1 & b & 0.112  &  vs11h & B &      &          & 05-06-96 &  \\
J0509$+$0541 & 0506$+$056 &   1.0 & e &        &  vs11w & B &      &          & 05-06-96 &  \\
J0530$+$1331 & 0528$+$134 &   6.2 & q & 2.060  &  vs01o & B &      &          & 05-06-96 & M\\
J0532$+$0732 & 0529$+$075 &   2.0 & e &        &        & C &      &          & 05-06-96 &  \\
J0539$-$2839 & 0537$-$286 &   1.2 & q & 3.104  &  vs10g & A &  1.4 & 01-10-99 & 05-06-96 &  \\
J0541$-$0541 & 0539$-$057 &   1.2 & q & 0.839  &  vs07y & B &      &          & 05-06-96 &  \\
J0607$+$6720 & 0602$+$673 &   1.0 & q &        &  vs11v & B &      &          & 05-06-96 & M\\
J0607$-$0834 & 0605$-$085 &   2.4 & q & 0.870  &  vs03p & A &  0.4 & 14-01-99 & 05-06-96 & M\\
J0609$-$1542 & 0607$-$157 &   4.2 & q & 0.324  &  vs02a & A &  0.9 & 04-04-98 & 05-06-96 &  \\
J0614$+$6046 & 0609$+$607 &   1.1 & q & 2.690  &  vs10f & B &      &          & 05-06-96 &  \\
J0616$-$3456 & 0614$-$349 &   1.4 & g & 0.329  &        & D &      &          & 05-06-96 &  \\
J0620$-$3711 & 0618$-$371 &   1.4 & g & 0.033  &        & D &      &          & 05-06-96 &  \\
J0626$+$8202 & 0615$+$820 &   1.0 & q & 0.710  &  vs11u & B &      &          & 18-09-99 & M\\
J0627$-$3529 & 0625$-$354 &   2.2 & g & 0.055  &        & C &      &          & 05-06-96 &  \\
J0644$-$3459 & 0642$-$349 &   1.0 & q & 2.165  &  vs10e & B &      &          & 05-06-96 &  \\
J0646$+$4451 & 0642$+$449 &   1.7 & q & 3.396  &  vs05l & B &      &          & 23-06-99 & M\\
J0648$-$3044 & 0636$-$306 &   1.0 & q & 0.455  &  vs11t & B &      &          & 05-06-96 &  \\
J0713$+$4349 & 0710$+$439 &   1.6 & q & 0.518  &  vs06p & B &      &          & 05-06-96 & M\\
J0714$+$3534 & 0711$+$356 &   1.0 & q & 1.620  &  vs09k & A &  1.4 & 09-04-99 & 09-04-99 &  \\
J0738$+$1742 & 0735$+$178 &   2.2 & b & 0.410  &  vs05d & A &  1.2 & 30-01-99 & 30-01-99 & M\\
J0739$+$0137 & 0736$+$017 &   1.8 & q & 0.191  &  vs05c & A &  2.1 & 03-04-99 & 05-06-96 & M\\
J0741$+$3112 & 0738$+$313 &   3.4 & q & 0.635  &  vs02k & A &  3.2 & 10-01-99 & 10-01-99 & M\\
J0745$+$1011 & 0742$+$103 &   3.5 & e &        &  vs02j & B &      &          & 05-06-96 & M\\
J0745$-$0044 & 0743$-$006 &   2.0 & b & 0.994  &  vs04f & B &      &          & 05-06-96 &  \\
J0748$+$2400 & 0745$+$241 &   1.2 & q & 0.410  &  vs10d & A &  0.6 & 09-02-99 & 05-06-96 & M\\
J0750$+$1231 & 0748$+$126 &   1.9 & q & 0.889  &  vs04k & B &      &          & 05-06-96 & M\\
J0758$+$3747 & NGC2484    &   1.0 & g & 0.043  &        & C &      &          & 05-06-96 &  \\
J0808$-$0751 & 0805$-$077 &   1.6 & q & 1.837  &        & C &      &          & 05-06-96 &  \\
J0808$+$4950 & 0804$+$499 &   1.2 & q & 1.430  &  vs09j & B &      &          & 05-06-96 & M\\
J0811$+$0146 & 0808$+$019 &   1.5 & b &        &  vs07n & A &  1.6 & 07-01-99 & 07-01-99 & M\\
J0815$+$3635 & 0812$+$367 &   1.0 & q & 1.025  &  vs10c & B &      &          & 05-06-96 &  \\
J0818$+$4222 & 0814$+$425 &   1.9 & b & 0.0257 &  vs06g & A &  1.7 & 24-04-99 & 24-04-99 & M\\
J0820$-$1258 & 0818$-$128 &   1.0 & b &        &  vs11s & B &      &          & 05-06-96 &  \\
J0823$+$2223 & 0820$+$225 &   1.6 & b & 0.951  &        & C &      &          & 05-06-96 &  \\
J0824$+$5552 & 0820$+$560 &   1.2 & q & 1.417  &  vs09i & A &  1.6 & 15-10-00 & 05-06-96 &  \\
J0824$+$3916 & 0821$+$394 &   1.0 & q & 1.216  &  vs10b & B &      &          & 05-06-96 &  \\
J0825$+$0309 & 0823$+$033 &   1.6 & b & 0.506  &  vs05v & A &  0.4 & 28-04-99 & 05-06-96 & M\\
J0831$+$0429 & 0829$+$046 &   2.1 & b & 0.180  &  vs05k & B &      &          & 05-06-96 & M\\
J0834$+$5534 & 0831$+$557 &   5.8 & g & 0.242  &        & C &      &          & 05-06-96 &  \\
J0840$+$1312 & 0838$+$133 &   1.3 & q & 0.684  &  vs09h & B &      &          & 05-06-96 &  \\
J0841$+$7053 & 0836$+$710 &   2.4 & q & 2.172  &  vs04e & A &  2.7 & 07-10-97 & 07-10-97 & M\\
J0842$+$1835 & 0839$+$187 &   1.0 & q & 1.270  &  vs10a & B &      &          & 05-06-96 &  \\
J0854$+$5757 & 0850$+$581 &   1.2 & q & 1.322  &  vs09z & B &      &          & 05-06-96 & M\\
J0854$+$2006 & OJ287      &   2.7 & b & 0.306  &  vs03y & A &  2.9 & 04-04-99 & 04-04-99 &  \\
J0900$-$2808 & 0858$-$279 &   3.6 & q & 2.152  &        & C &      &          & 05-06-96 &  \\
J0903$+$4651 & 0859$+$470 &   1.3 & q & 1.462  &  vs09g & A &  1.1 & 14-02-99 & 14-02-99 &  \\
J0909$+$0121 & 0906$+$015 &   1.0 & q & 1.018  &  vs11r & A &  1.0 & 09-01-99 & 05-06-96 & M\\
J0920$+$4441 & 0917$+$449 &   1.2 & q & 2.180  &  vs07x & A &  3.0 & 07-02-99 & 07-02-99 & M\\
J0921$-$2618 & 0919$-$260 &   2.3 & q & 2.300  &  vs04s & B &      &          & 05-06-96 &  \\
J0921$+$6215 & 0917$+$624 &   1.5 & q & 1.446  &  vs06f & B &      &          & 05-06-96 &  \\
J0927$+$3902 & 4C39.25    &  11.2 & q & 0.699  &  vs01f & A &  2.6 & 23-10-97 & 23-10-97 & M\\
J0948$+$4039 & 0945$+$408 &   1.6 & q & 1.252  &  vs05u & B &      &          & 05-06-96 & M\\
J0956$+$2515 & 0953$+$254 &   1.3 & q & 0.712  &  vs07l & B &      &          & 05-06-96 & M\\
J0957$+$5522 & 0954$+$556 &   2.3 & q & 0.909  &        & D &      &          & 05-06-96 &  \\
J0958$+$4725 & 0955$+$476 &   1.3 & q & 1.873  &  vs07k & B &      &          & 05-06-96 &  \\
J0958$+$6533 & 0954$+$658 &   1.4 & b & 0.368  &  vs08k & B &      &          & 05-06-96 &  \\
J1006$+$3454 & 3C236      &   1.7 & g & 0.0989 &        & D &      &          & 05-06-96 &  \\
J1014$+$2301 & 1012$+$232 &   1.1 & q & 0.565  &  vs11g & B &      &          & 05-06-96 & M\\
J1018$-$3144 & 1015$-$314 &   1.8 & q & 1.346  &        & D &      &          & 05-06-96 &  \\
J1035$-$2011 & 1032$-$199 &   1.1 & q & 2.189  &  vs11f & A &  1.7 & 21-02-98 & 05-06-96 & M\\
J1035$+$5628 & 1031$+$567 &   1.3 & g & 0.450  &        & C &      &          & 05-06-96 &  \\
J1037$-$2934 & 1034$-$293 &   1.6 & b & 0.312  &  vs05t & A &  2.8 & 02-06-99 & 02-06-99 &  \\
J1041$+$0610 & 1038$+$064 &   1.7 & q & 1.270  &  vs05j & B &      &          & 05-06-96 &  \\
J1042$+$1203 & 3C245      &   1.7 & q & 1.029  &        & C &      &          & 05-06-96 &  \\
J1044$+$8054 & 1039$+$811 &   1.1 & q & 1.260  &  vs11e & B &      &          & 05-06-96 &  \\
J1048$-$1909 & 1045$-$188 &   1.1 & q & 0.595  &  vs11d & B &      &          & 05-06-96 &  \\
J1048$-$4114 & 1045$-$188 &   1.0 & q & 0.620  &        & C &      &          & 05-06-96 &  \\
J1051$-$3138 & 1048$-$313 &   1.0 & e &        &  vs09y & A &  0.1 & 20-05-99 & 05-06-96 &  \\
J1051$+$2119 & 1049$+$215 &   1.3 & q & 1.300  &  vs07j & B &      &          & 05-06-96 & M\\
J1058$+$0133 & 1055$+$018 &   4.1 & b & 0.888  &  vs02l & A &  7.3 & 12-05-99 & 12-05-99 & M\\
J1058$+$1951 & 1055$+$201 &   1.7 & q & 1.110  &  vs06x & B &      &          & 05-06-96 & M\\
J1118$+$1234 & 1116$+$128 &   2.0 & q & 2.118  &  vs05s & A &  1.0 & 17-12-97 & 05-06-96 &  \\
J1125$+$2610 & 1123$+$264 &   1.1 & q & 2.341  &  vs09f & B &      &          & 05-06-96 & M\\
J1127$-$1857 & 1124$-$186 &   1.6 & q & 1.050  &  vs07i & B &      &          & 05-06-96 & M\\
J1146$-$2447 & 1143$-$245 &   1.5 & q & 1.950  &  vs06w & A &  4.2 & 26-05-99 & 05-06-96 &  \\
J1146$-$3328 & 1143$-$331 &   1.1 & g &        &        & C &      &          & 05-06-96 &  \\
J1147$-$3812 & 1144$-$379 &   2.2 & b & 1.048  &  vs04d & A &  0.2 & 28-12-97 & 05-06-96 &  \\
J1147$-$0724 & 1145$-$071 &   1.2 & q & 1.342  &  vs09x & A &  1.5 & 08-03-00 & 05-06-96 & M\\
J1150$-$0023 & 1148$-$001 &   2.0 & q & 1.976  &  vs05r & B &      &          & 05-06-96 & M\\
J1153$+$4931 & 1150$+$497 &   1.0 & q & 0.334  &  vs09w & B &      &          & 05-06-96 &  \\
J1153$+$8058 & 1150$+$812 &   1.4 & q & 1.250  &  vs06v & B &      &          & 05-06-96 &  \\
J1158$+$2450 & 1155$+$251 &   1.2 & g &        &        & C &      &          & 05-06-96 & M\\
J1159$+$2914 & 1156$+$295 &   1.8 & q & 0.729  &  vs05b & B &      &          & 05-06-96 & M\\
J1205$-$2634 & 1203$-$262 &   1.1 & q & 0.789  &  vs11c & B &      &          & 05-06-96 &  \\
J1209$-$2406 & 1206$-$238 &   1.1 & q &        &  vs11b & B &      &          & 05-06-96 &  \\
J1215$-$1731 & 1213$-$172 &   1.8 & g &        &  vs05a & A &  0.9 & 11-01-98 & 05-06-96 &  \\
J1215$+$3448 & 1213$+$350 &   1.1 & q & 0.857  &        & C &      &          & 05-06-96 &  \\
J1219$+$4829 & 1216$+$487 &   1.0 & q & 1.076  &  vs11a & B &      &          & 05-06-96 &  \\
J1224$+$0330 & 1222$+$037 &   1.2 & q & 0.957  &  vs09e & B &      &          & 05-06-96 &  \\
J1224$+$2122 & 1222$+$216 &   1.4 & q & 0.435  &  vs06o & B &      &          & 05-06-96 &  \\
J1225$+$1253 & 3C272.1    &   3.6 & g & 0.0033 &        & D &      &          & 05-06-96 &  \\
J1229$+$0203 & 3C273B     &  43.6 & q & 0.1583 &  vs01b & A &  1.4 & 22-12-97 & 22-12-97 & M\\
J1232$-$0224 & 1229$-$021 &   1.0 & q & 1.038  &        & C &      &          & 05-06-96 &  \\
J1239$-$1023 & 1237$-$101 &   1.3 & q & 0.750  &        & C &      &          & 05-06-96 &  \\
J1246$-$0730 & 1243$-$072 &   1.1 & q & 1.286  &  vs10z & B &      &          & 05-06-96 &  \\
J1246$-$2547 & 1244$-$255 &   2.3 & q & 0.638  &  vs04r & A &  6.5 & 21-01-98 & 05-06-96 &  \\
J1256$-$0547 & 3C279      &  13.0 & q & 0.538  &  vs01g & A &  4.6 & 10-01-98 & 10-01-98 & M\\
J1257$-$3155 & 1255$-$316 &   1.4 & q & 1.924  &  vs06u & A &  2.1 & 25-05-99 & 05-06-96 &  \\
J1305$-$1033 & 1302$-$102 &   1.0 & q & 0.286  &  vs11q & A &  1.1 & 07-01-98 & 05-06-96 & M\\
J1309$+$1154 & 1307$+$121 &   1.3 & b &        &  vs09d & B &      &          & 05-06-96 &  \\
J1310$+$3220 & 1308$+$326 &   3.6 & b & 0.996  &  vs02e & A & 10.0 & 29-06-98 & 29-06-98 & M\\
J1316$-$3338 & 1313$-$333 &   1.6 & q & 1.210  &  vs05q & A &  0.6 & 20-01-98 & 05-06-96 &  \\
J1332$+$0200 & 1330$+$022 &   1.5 & g & 0.215  &        & C &      &          & 05-06-96 &  \\
J1337$-$1257 & 1334$-$127 &   4.4 & b & 0.539  &  vs01y & A &  1.0 & 10-07-99 & 10-07-99 & M\\
J1347$+$1217 & 1345$+$125 &   3.1 & g & 0.1202 &        & C &      &          & 05-06-96 & M\\
J1351$-$1449 & 1349$-$145 &   1.1 & e &        &  vs10y & B &      &          & 05-06-96 &  \\
J1354$-$1041 & 1352$-$104 &   1.0 & q & 0.332  &        & C &      &          & 05-06-96 &  \\
J1357$+$1919 & 1354$+$195 &   2.7 & q & 0.720  &  vs03x & A &  0.3 & 19-07-99 & 05-06-96 &  \\
J1357$-$1744 & 1354$-$174 &   1.1 & q & 3.147  &  vs08j & A &  1.1 & 30-01-98 & 05-06-96 &  \\
J1405$+$0415 & 1402$+$044 &   1.0 & q & 3.193  &  vs10x & B &      &          & 05-06-96 &  \\
J1407$+$2827 & OQ208      &   2.4 & g & 0.0769 &  vs03o & A &  3.1 & 30-06-98 & 30-06-98 & M\\
J1408$-$0752 & 1406$-$076 &   1.0 & q & 1.494  &  vs11p & B &      &          & 05-06-96 &  \\
J1415$+$1320 & 1413$+$135 &   1.2 & b & 0.260  &  vs09v & B &      &          & 05-06-96 & M\\
J1419$+$5423 & 1418$+$546 &   1.7 & b & 0.152  &        & C &      &          & 05-06-96 &  \\
J1419$-$1928 & 1417$-$192 &   1.0 & g & 0.119  &  vs11o & B &      &          & 05-06-96 &  \\
J1427$-$4206 & 1424$-$418 &   3.8 & q & 1.522  &  vs02i & B &      &          & 05-06-96 &  \\
J1430$+$1043 & 1427$+$109 &   1.2 & q & 1.710  &  vs09c & A &  0.6 & 18-01-01 & 05-06-96 &  \\
J1436$+$6336 & 1435$+$638 &   1.1 & q & 2.068  &  vs08i & A &  1.1 & 21-06-01 & 05-06-96 &  \\
J1445$+$0958 & 1442$+$101 &   1.2 & q & 3.535  &        & C &      &          & 05-06-96 &  \\
J1454$-$3747 & 1451$-$375 &   2.4 & q & 0.314  &  vs03n & B &      &          & 05-06-96 &  \\
J1501$-$3918 & 1458$-$391 &   1.2 & g & 1.083  &        & C &      &          & 05-06-96 &  \\
J1504$+$1029 & 1502$+$106 &   1.8 & q & 1.839  &  vs04z & A &  2.5 & 27-07-00 & 05-06-96 & M\\
J1506$+$3730 & 1504$+$377 &   1.0 & g & 0.674  &  vs10w & B &      &          & 05-06-96 & M\\
J1507$-$1652 & 1504$-$166 &   2.8 & q & 0.876  &  vs03m & A &  0.5 & 10-04-98 & 05-06-96 & M\\
J1512$-$0905 & 1510$-$089 &   3.3 & q & 0.360  &  vs02x & A &  4.8 & 11-08-99 & 11-08-99 & M\\
J1513$-$1012 & 1511$-$100 &   1.2 & q & 1.513  &  vs09u & B &      &          & 05-06-96 & M\\
J1516$+$0015 & 1514$+$004 &   1.6 & g & 0.0523 &  vs07h & A &  1.3 & 24-08-98 & 05-06-96 &  \\
J1517$-$2422 & 1514$-$241 &   2.2 & g & 0.042  &  vs03u & A &  1.2 & 27-04-98 & 05-06-96 & M\\
J1522$-$2730 & 1519$-$273 &   2.3 & b & 1.294  &  vs04q & A &  0.8 & 05-02-99 & 05-02-99 & M\\
J1526$-$1351 & 1524$-$136 &   1.3 & q & 1.687  &        & C &      &          & 05-06-96 &  \\
J1534$+$0131 & 1532$+$016 &   1.3 & q & 1.435  &        & C &      &          & 05-06-96 & M\\
J1540$+$1447 & 1538$+$149 &   1.2 & b & 0.605  &  vs09t & B &      &          & 05-06-96 &  \\
J1543$-$0757 & 1540$-$077 &   1.0 & g &        &        & C &      &          & 05-06-96 &  \\
J1546$+$0026 & 1543$+$005 &   1.3 & g & 0.550  &  vs09b & B &      &          & 05-06-96 &  \\
J1549$+$0237 & 1546$+$027 &   1.5 & q & 0.413  &  vs06e & A &  2.4 & 31-07-98 & 05-06-96 & M\\
J1550$+$0527 & 1548$+$056 &   3.3 & q & 1.422  &  vs02r & B &      &          & 05-06-96 & M\\
J1557$-$0001 & 1555$+$001 &   2.3 & q & 1.770  &  vs04p & B &      &          & 05-06-96 &  \\
J1602$+$3326 & 1600$+$335 &   2.0 & g &        &  vs04y & B &      &          & 05-06-96 &  \\
J1608$+$1029 & 1606$+$106 &   1.7 & q & 1.226  &  vs06s & B &      &          & 05-06-96 & M\\
J1613$+$3412 & 1611$+$343 &   4.0 & q & 1.401  &  vs02b & A &  1.1 & 04-02-98 & 05-06-96 & M\\
J1625$-$2527 & 1622$-$253 &   3.5 & g & 0.786  &  vs02h & B &      &          & 05-06-96 & M\\
J1625$+$4134 & 1624$+$416 &   1.4 & q & 2.550  &        & C &      &          & 05-06-96 &  \\
J1626$-$2951 & 1622$-$297 &   2.4 & q & 0.815  &  vs03l & A &  0.6 & 22-02-98 & 05-06-96 &  \\
J1635$+$3808 & 1633$+$382 &   3.2 & q & 1.814  &  vs03d & A &  0.6 & 04-08-98 & 04-08-98 & M\\
J1637$+$4717 & 1636$+$473 &   1.3 & q & 0.740  &  vs09a & B &      &          & 05-06-96 &  \\
J1638$+$5720 & 1637$+$574 &   1.8 & q & 0.745  &  vs06n & A &  0.7 & 21-04-98 & 21-04-98 &  \\
J1640$+$3946 & NRAO512    &   1.3 & q & 1.660  &  vs08y & A &  0.3 & 03-09-99 & 05-06-96 & M\\
J1640$+$1220 & 1638$+$124 &   1.3 & g &        &  vs08z & B &      &          & 05-06-96 &  \\
J1642$+$3948 & 3C345      &   8.4 & q & 0.593  &  vs01k & A &  5.5 & 28-07-98 & 28-07-98 & M\\
J1642$+$6856 & 1642$+$690 &   1.5 & q & 0.751  &  vs07w & A &  0.8 & 31-05-98 & 31-05-98 & M\\
J1642$-$0621 & 1639$-$062 &   1.6 & e &        &  vs13m & B &      &          & 05-06-96 & M\\
J1653$+$3945 & MARK501    &   1.4 & b & 0.0336 &  vs08h & A &  0.7 & 07-04-98 & 07-04-98 &  \\
J1658$+$4737 & 1656$+$477 &   1.4 & q & 1.622  &  vs08g & A &  0.2 & 28-08-99 & 05-06-96 &  \\
J1658$+$0741 & 1655$+$077 &   1.6 & q & 0.621  &  vs07g & A &  1.7 & 05-03-98 & 05-06-96 & M\\
J1658$+$0515 & 1656$+$053 &   2.1 & q & 0.879  &  vs05i & A &  1.2 & 04-03-98 & 05-06-96 & M\\
J1658$-$0739 & 1656$-$075 &   1.3 & q & 3.700  &        & C &      &          & 05-06-96 &  \\
J1710$+$0036 & 1708$+$006 &   1.1 & g &        &        & D &      &          & 05-06-96 &  \\
J1727$+$4530 & 1726$+$455 &   1.3 & q & 0.714  &  vs07f & B &      &          & 05-06-96 &  \\
J1728$+$0427 & 1725$+$044 &   1.2 & q & 0.296  &        & C &      &          & 05-06-96 &  \\
J1733$-$1304 & NRAO530    &   7.0 & q & 0.902  &  vs01m & A &  2.6 & 08-09-97 & 05-06-96 & M\\
J1734$+$3857 & 1732$+$389 &   1.3 & b & 0.970  &  vs07e & B &      &          & 05-06-96 &  \\
J1740$+$5211 & 1739$+$522 &   1.1 & q & 1.375  &  vs10v & A &  4.3 & 14-06-98 & 14-06-98 & M\\
J1743$-$0350 & 1741$-$038 &   2.4 & q & 1.054  &  vs03k & B &      &          & 05-06-96 & M\\
J1745$-$0753 & 1742$-$078 &   1.4 & e &        &        & C &      &          & 05-06-96 &  \\
J1751$+$0939 & 1749$+$096 &   2.3 & b & 0.322  &  vs04o & A &  0.7 & 20-08-98 & 20-08-98 & M\\
J1753$+$4409 & 1751$+$441 &   1.0 & q & 0.871  &  vs11n & B &      &          & 05-06-96 &  \\
J1800$+$7828 & 1803$+$784 &   2.6 & b & 0.680  &  vs02w & A &  0.9 & 16-10-97 & 16-10-97 & M\\
J1801$+$4404 & 1800$+$440 &   1.1 & q & 0.663  &  vs10u & B &      &          & 05-06-96 &  \\
J1804$+$0101 & 1801$+$010 &   1.6 & q & 1.522  &        & C &      &          & 05-06-96 &  \\
J1806$+$6949 & 3C371      &   2.2 & b & 0.051  &  vs04x & A &  5.3 & 11-03-98 & 11-03-98 &  \\
J1824$+$1044 & 1821$+$107 &   1.1 & q & 1.360  &  vs10t & B &      &          & 05-06-96 &  \\
J1824$+$5651 & 1823$+$568 &   1.5 & b & 0.664  &  vs06d & A &  0.6 & 31-05-98 & 31-05-98 & M\\
J1832$+$2833 & 1830$+$285 &   1.1 & q & 0.594  &  vs10s & B &      &          & 05-06-96 &  \\
J1902$+$3159 & 3C395      &   1.9 & q & 0.635  &  vs06c & A &  1.2 & 01-05-98 & 01-05-98 & M\\
J1911$-$2006 & 1908$-$201 &   2.3 & e &        &  vs04n & B &      &          & 05-06-96 &  \\
J1924$-$2914 & 1921$-$293 &  14.8 & q & 0.352  &  vs01e & A &  1.4 & 19-06-98 & 19-06-98 & M\\
J1927$+$7358 & 1928$+$738 &   3.6 & q & 0.303  &  vs02m & A &  2.1 & 22-08-97 & 05-06-96 & M\\
J1937$-$3958 & 1933$-$400 &   1.1 & q & 0.966  &  vs08f & B &      &          & 05-06-96 &  \\
J1949$-$1957 & 1946$-$200 &   1.3 & e &        &  vs13c & B &      &          & 05-06-96 &  \\
J1955$+$5131 & 1954$+$513 &   3.1 & q & 1.220  &  vs13d & B &      &          & 05-06-96 & M\\
J1957$-$3845 & 1954$-$388 &   4.4 & q & 0.630  &  vs01x & B &      &          & 05-06-96 &  \\
J2000$-$1748 & 1958$-$179 &   1.8 & q & 0.650  &  vs04w & A &  2.4 & 25-06-98 & 05-06-96 &  \\
J2003$-$3251 & 2000$-$330 &   1.2 & q & 3.773  &  vs08x & B &      &          & 05-06-96 &  \\
J2005$+$7752 & 2007$+$777 &   1.7 & b & 0.342  &  vs06r & A &  1.0 & 10-03-98 & 10-03-98 & M\\
J2011$-$1546 & 2008$-$159 &   1.4 & q & 1.180  &  vs07d & A &  0.8 & 25-10-97 & 05-06-96 &  \\
J2022$+$6136 & 2021$+$614 &   3.0 & q & 0.227  &  vs02q & A &  1.1 & 06-11-97 & 06-11-97 & M\\
J2031$+$1219 & 2029$+$121 &   1.2 & b & 1.215  &  vs08w & B &      &          & 05-06-96 &  \\
J2101$+$0341 & 2059$+$034 &   1.3 & q & 1.105  &  vs08v & A &  0.9 & 14-11-00 & 05-06-96 &  \\
J2109$-$4110 & 2106$-$413 &   2.3 & q & 1.055  &  vs03w & B &      &          & 05-06-96 &  \\
J2110$-$1020 & 2107$-$105 &   1.2 & q &        &  vs09r & B &      &          & 05-06-96 &  \\
J2115$+$2933 & 2113$+$293 &   1.2 & q & 1.514  &  vs09q & B &      &          & 05-06-96 & M\\
J2123$+$0535 & 2121$+$053 &   3.2 & q & 1.878  &  vs03c & B &      &          & 05-06-96 & M\\
J2129$-$1538 & 2126$-$158 &   1.2 & q & 3.268  &  vs08e & A &  1.6 & 13-07-98 & 13-07-98 &  \\
J2131$-$1207 & 2128$-$123 &   3.0 & q & 0.501  &  vs13e & B &      &          & 05-06-96 & M\\
J2136$+$0041 & 2134$+$004 &  12.4 & q & 1.392  &  vs01h & A &  1.3 & 28-11-97 & 05-06-96 & M\\
J2139$+$1423 & 2136$+$141 &   2.0 & q & 2.427  &  vs04c & A &  2.9 & 25-08-98 & 05-06-96 & M\\
J2147$+$0929 & 2144$+$092 &   1.1 & q & 1.113  &  vs10r & B &      &          & 05-06-96 & M\\
J2148$+$0657 & 2145$+$067 &   6.4 & q & 0.990  &  vs01l & B &      &          & 05-06-96 & M\\
J2151$+$0552 & 2149$+$056 &   1.0 & g & 0.740  &  vs10q & B &      &          & 05-06-96 &  \\
J2151$-$3027 & 2149$-$306 &   1.4 & q & 2.345  &        & C &      &          & 05-06-96 &  \\
J2158$-$1501 & 2155$-$152 &   3.1 & b & 0.672  &  vs03g & A &  1.2 & 15-08-98 & 05-06-96 & M\\
J2202$+$4216 & BLLAC      &   5.6 & b & 0.0686 &  vs01q & A &  2.8 & 08-12-97 & 08-12-97 & M\\
J2203$+$3145 & 2201$+$315 &   2.9 & q & 0.301  &  vs13f & B &      &          & 05-06-96 & M\\
J2206$-$1835 & 2203$-$188 &   4.3 & q & 0.619  &        & D &      &          & 05-06-96 &  \\
J2212$+$2355 & 2209$+$236 &   1.4 & q &        &  vs06l & A &  2.3 & 27-05-98 & 05-06-96 & M\\
J2213$-$2529 & 2210$-$257 &   1.0 & q & 1.833  &  vs13h & B &      &          & 05-06-96 &  \\
J2218$-$0335 & 2216$-$038 &   2.7 & q & 0.901  &  vs03j & A &  0.9 & 03-12-97 & 05-06-96 &  \\
J2225$+$2118 & 2223$+$210 &   1.2 & q & 1.959  &        & C &      &          & 05-06-96 &  \\
J2225$-$0457 & 3C446      &   6.4 & b & 1.404  &  vs01s & A &  1.6 & 02-12-00 & 05-06-96 & M\\
J2229$-$0832 & 2227$-$088 &   2.4 & q & 1.561  &  vs04i & A &  1.3 & 27-11-97 & 05-06-96 & M\\
J2230$+$6946 & 2229$+$695 &   1.4 & g &        &        & C &      &          & 05-06-96 & M\\
J2236$+$2828 & 2234$+$282 &   1.6 & q & 0.795  &  vs07c & A &  0.3 & 24-07-99 & 05-06-96 &  \\
J2243$-$2544 & 2240$-$260 &   1.2 & b & 0.774  &        & C &      &          & 05-06-96 &  \\
J2246$-$1206 & 2243$-$123 &   2.7 & q & 0.630  &  vs03v & A &  2.3 & 03-06-98 & 05-06-96 & M\\
J2249$+$1136 & NGC7385    &   1.1 & g & 0.026  &        & D &      &          & 05-06-96 &  \\
J2250$+$1419 & 2247$+$140 &   1.2 & q & 0.237  &        & D &      &          & 05-06-96 &  \\
J2253$+$1608 & 3C454.3    &  16.0 & q & 0.859  &  vs01d & A &  0.7 & 12-12-97 & 12-12-97 & M\\
J2255$+$4202 & 2253$+$417 &   1.1 & q & 1.476  &  vs13g & B &      &          & 05-06-96 &  \\
J2258$-$2758 & 2255$-$282 &   2.5 & q & 0.926  &  vs03f & B &      &          & 05-06-96 & M\\
J2311$+$3425 & 2308$+$341 &   1.0 & q & 1.817  &  vs13n & B &      &          & 05-06-96 &  \\
J2320$+$0513 & 2318$+$049 &   1.2 & q & 0.623  &  vs09p & A &  1.3 & 08-12-00 & 05-06-96 & M\\
J2321$+$2732 & 2319$+$272 &   1.0 & q & 1.253  &  vs13i & B &      &          & 05-06-96 &  \\
J2330$+$1100 & 2328$+$107 &   1.2 & q & 1.489  &  vs13j & B &      &          & 05-06-96 &  \\
J2331$-$1556 & 2329$-$162 &   1.9 & q & 1.155  &  vs06b & A &  0.5 & 11-06-98 & 05-06-96 &  \\
J2333$-$2343 & 2331$-$240 &   1.1 & g & 0.0477 &        & C &      &          & 05-06-96 &  \\
J2346$+$0930 & 2344$+$092 &   1.4 & q & 0.677  &  vs13k & B &      &          & 05-06-96 &  \\
J2354$+$4553 & 2351$+$456 &   1.2 & q & 1.992  &  vs13l & B &      &          & 05-06-96 &  \\
J2357$-$1125 & 2354$-$116 &   1.4 & q & 0.960  &  vs13l & B &      &          & 05-06-96 &  \\
J2358$-$1020 & 2355$-$106 &   1.6 & q & 1.622  &  vs06k & A &  0.4 & 10-12-00 & 05-06-96 &  \\

\enddata \footnotesize \tablecomments {Column (7): A=AGN sample and
observed with VSOP; B=AGN sample, but not yet observed with VSOP or reduced;
C=AGN sample, but too faint to be observed with VSOP; D=Removed from
AGN sample.
Redshift and identification for J1501$-$3918 and J1658$-$0739 from I.A.G. Snellin \& P.G. Edwards
(2004, private communication); redshift for J1522$-$2730 from \cite{hei04}}
\end{deluxetable}
}}
\end{document}